\documentclass[final,5p,times,sort,compress]{elsarticle}
\graphicspath{{./figs/}}
\usepackage{changes}
\usepackage[english]{babel}
\usepackage{microtype}
\usepackage[all]{nowidow}
\usepackage[colorlinks=true, linkcolor=blue, citecolor=blue, urlcolor=blue]{hyperref} 
\usepackage{amsmath,amssymb}
\journal{Powder Technology}
\usepackage{booktabs}
\usepackage{siunitx} 
\usepackage{verbatim}
\usepackage{hyperref}
\usepackage{cleveref}
\Crefname{equation}{Equation}{Equations}
\Crefname{equation}{Equation}{Equations}
\Crefname{figure}{Figure}{Figures}
\Crefname{figure}{Figure}{Figures}
\Crefname{section}{Section}{Sections}
\Crefname{subsection}{Section}{Sections}
\Crefname{table}{Table}{Tables}
\Crefname{appendix}{}{}
\usepackage{enumitem}
\usepackage{float}

\usepackage{tikz}
\usetikzlibrary{positioning,shapes,arrows,arrows.meta}

\tikzstyle{startstop} = [rectangle, rounded corners, 
minimum width=3cm, 
minimum height=0.5cm,
text centered, 
draw=black, 
fill=red!30]
\tikzstyle{io} = [trapezium, 
trapezium stretches=true, 
trapezium left angle=70, 
trapezium right angle=110, 
minimum width=3cm, 
minimum height=1cm, text centered, 
draw=black, fill=blue!30]
\tikzstyle{process} = [rectangle, 
minimum width=3cm, 
minimum height=0.5cm, 
text centered, 
text width=3.5cm, 
draw=black, 
fill=orange!30]

\tikzstyle{decision} = [diamond, 
minimum width=3cm, 
minimum height=1cm, 
text centered, 
draw=black, 
fill=green!30]
\tikzstyle{arrow} = [-{Stealth[scale=1.2]},rounded corners,thick]

\usepackage{stmaryrd}
\newcommand{\truemax}{\varoast}
\newcommand{\voxmax}{\ast}
\newcommand{\latconst}{h}

\begin{document}
\begin{frontmatter}

\title{Multi-sphere shape generator for DEM simulations of complex-shaped particles}
\author{Felix Buchele\corref{corauth}}
\cortext[cortext]{Corresponding author}
\ead{felix.buchele@fau.de}
\author{Thorsten P\"oschel}
\author{Patric M\"uller}
\address{Institute for Multiscale Simulation, Friedrich-Alexander-Universit\"at Erlangen-N\"urnberg, Erlangen, GERMANY}

\begin{abstract}
MSS is an algorithm to determine the radii and positions of spheres that fill a given volume. In the context of granular materials, MSS is a particle generator for DEM simulations of complex-shaped particles. Here, each particle of a given shape is represented by a set of spheres that collectively approximate the particle. This technique of particle shape representation is often referred to as the multi-sphere approach. We show that, for a given number of spheres, MSS provides a closer approximation to the target shape at lower computational costs than other DEM multi-sphere particle generators reported in the literature.
\end{abstract}


\end{frontmatter}


\section{Introduction}

In many cases, the mechanical and rheological properties of particulate systems depend strongly on the shape of the individual particles. Reliable simulations, therefore, take the particle shape into account. In discrete-element simulations (DEM) \cite{Cundall:1979, Poeschel.2005, Matuttis:2014}, the multi-sphere approach is the most widely used method for representing non-spherical particles. Here, each individual particle is represented by a set of spheres. The multi-sphere approach was first used to simulate avalanches and stick-slip flow of granular matter \cite{Poeschel.1993, BuchholtzPoeschel:1994, BuchholtzPoeschel:1996} and later generalized  to a broader range of applications \cite{Hubbard:1996,Favier:1999, Favier.2001}. A major advantage of the multi-sphere approach over other models is that contacts between complex-shaped particles can be reduced to interactions between the constituent spheres \cite{Berry:2023}. A limitation of this model is that representing sharp-edged particles requires a large number of spheres. Complex-shaped particles can consist of hundreds or even thousands of spheres. Since the number of spheres involved has a significant impact on the computational cost, it is crucial to represent a given particle shape with the required precision, using as few spheres as possible. Consequently, considerable effort has been dedicated to solving this problem, namely, determining the radii and relative positions of the constituent spheres that best represent a complex-shaped particle \cite{Favier:1999, Bradshaw.2004, Price.2007, Wang.2007, Cho.2007, Garcia.2009, Markauskas.2010, Ferellec.2010, Taghavi.2011, Amberger:2012, Gao.2012, Li.2015, Pasha.2016, Zheng.2016, Haeri.2017, Nan.2017, Katagiri.2019, Yuan:2019, Tamadondar.2020, Kodicherla.2020, Angelidakis.2021}. A critical review and a quantitative comparison of these approaches can be found in Ref. \cite{Angelidakis.2021}.

In the current paper, we introduce MSS (\underline{M}ulti-\underline{S}phere \underline{S}hape), a particle generator for DEM-particles of arbitrary geometric shape. We will demonstrate that our algorithm achieves higher accuracy than the aforementioned references when using the same number of spheres. MSS better preserves smooth surfaces and the symmetry of the particle shape, at lower computational effort.

\section{Problem description}
\subsection{Optimal sets of spheres}
The purpose of a multi-sphere particle generator is to approximate a given particle shape, $\cal S$,  by a set of spheres such that the shape is represented as accurately as possible. Each sphere, $i=0,\dots,n-1$, is specified by its center position $\vec{r}_i$ and its radius $R_i$, and occupies the space ${\cal \tilde{S}}_i\left(\vec{r}_i, R_i\right)$. 

The space occupied by the union of all spheres,
\begin{equation}
   {\cal \tilde{S}} \equiv \bigcup_{i=0}^{n-1} {\cal \tilde{S}}_i(R_i, \vec{r}_i)\,, 
   \label{eq:sphereRep}
\end{equation}
is referred to as the \textit{multi-sphere representation} of ${\cal S}$.

There are two alternative optimization problems:
\begin{itemize}
\item[(i)]  For a given number of spheres, $n$, determine their center locations 
and radii 
such that ${\cal \tilde{S}}$ provides the best approximation of ${\cal S}$ with respect to the volume of the symmetric difference,
\begin{equation}
    \min_{\left\{\vec{r}_i, R_i\right\}_{i=0}^{n-1}} \left|{\cal S}\triangle {\cal \tilde{S}}\right|
    \equiv 
    \min_{\left\{\vec{r}_i, R_i\right\}_{i=0}^{n-1}} \left|
\left( {\cal S} \cup {\cal \tilde{S}}\right) \setminus \left({\cal S}\cap {\cal \tilde{S}}\right) 
\right|    \,.
    \label{eq:mismatch-def}
\end{equation}

\item[(ii)] Determine the minimal number of spheres, $n$, their center locations and their radii such that the volume of the symmetric difference satisfies a specified quality criterion,
\begin{equation}
    n \to \min \quad \text{for}\quad  \min_{\left\{\vec{r}_i, R_i\right\}_{i=0}^{n-1}} \left|{\cal S}\triangle {\cal \tilde{S}}\right|<\varepsilon\,.
\label{eq:opt2}
\end{equation}

The quality criterion in \Cref{eq:opt2} can also be defined relative to the volume of the shape, 
\begin{equation}
    n \to \min \quad \text{for}\quad  \frac{ \min\limits_{\left\{\vec{r}_i, R_i\right\}_{i=0}^{n-1}} \left|{\cal S}\triangle {\cal \tilde{S}}\right|}
    {\left|{\cal S}\right|}
    < \eta\,.
\label{eq:opt2a}
\end{equation}
\end{itemize}

\subsection{Representation of the target particle shape, $S$}
\label{sec:voxel}
\label{sec:target-particle-shape}
In most DEM applications, particle shapes are either generated through rational design based on model assumptions, or derived from physically existing particles using 3D scanning or tomographic methods such as CT (Computed Tomography) and MRI (Magnetic Resonance Imaging). Shapes obtained through 3D scanning or rational design are typically represented as triangulated surface meshes. Shapes obtained from tomographic methods are usually provided as voxelized data.

MSS requires the representation of the shape ${\cal S}$ in binary voxel form, i.e., as a three-dimensional binary array
\begin{equation}
   S \quad\text{with}\quad  S_{ijk}\in\{0,1\}\,;~~~
   \begin{cases}
       0\le i \le n_x-1\\
       0\le j \le n_y-1\\
       0\le k \le n_z-1\,.
   \end{cases}
    \label{eq:voxel_S}
\end{equation}

We obtain $S$ by embedding the shape ${\cal S}$ into a three-dimen\-sional grid with lattice constant $\latconst$; $S_{ijk}=1$ if the grid cell $ijk$ is covered by more than 50\% by ${\cal S}$. Otherwise $S_{ijk}=0$. The parameters $\latconst$ and $n_x$, $n_y$, $n_z$ must be chosen such that 
\begin{equation}
\begin{array}{ll}
S_{0,j,k} = S_{1,j,k} = S_{n_{x}-2,j,k} = S_{n_{x}-1,j,k} = 0 \,;   & 0\le i \le n_x-1\\
S_{i,0,k} = S_{i,1,k} = S_{i,n_{y}-2,k} = S_{i,n_{y}-1,k} = 0  \,;   & 0\le j \le n_y-1\\
S_{i,j,0} = S_{i,j,1} = S_{i,j,n_{z}-2} = S_{i,j,n_{z}-1} = 0  \,;   & 0\le k \le n_z-1\,,
\end{array}
 \label{eq:S-rand}
\end{equation}
that is, the voxelized image $S$ of ${\cal S}$ is enclosed by a double-layered hull of array elements $S_{ijk}=0$.

Note that $\vec{r}_i$, $R_i$, ${\cal S}$, ${\cal \tilde{S}}_i$, and ${\cal \tilde{S}}$ are quantities defined in physical space, whereas $S$ and all other (later defined) voxel arrays are dimensionless. The transition between dimensional and dimensionless quantities is given by the lattice constant $\latconst$.

\subsection{Mismatch between the target shape and the multi-sphere representation}
\label{sec:mismatch}

In voxelized (dimensionless) representation, the mismatch $\left|{\cal S}\triangle {\cal \tilde{S}}\right|$ defined in \Cref{eq:mismatch-def} can be quantified by the relative difference
\begin{equation}
    \eta\left(S,\tilde{S}\right) \equiv\frac{\sum\limits_{i,j,k}\left|S_{ijk}-\tilde{S}_{ijk}\right|}{\sum\limits_{i,j,k} S_{ijk}}\,,
    \label{eq:mismatch-discrete}
\end{equation}
with the voxelized representation $\tilde{S}$ of ${\cal \tilde{S}}$ defined in the same way as the voxelized representation $S$ of ${\cal S}$, \Cref{eq:voxel_S}.  

A perfect but trivial optimum, $\eta\left(S,\tilde{S}\right)=0$, is obtained for any shape ${\cal S}$ if we place $n=\sum\limits_{ijk} S_{ijk}$ spheres with a (voxelized) radius $1/2$ at the center of each occupied voxel ($S_{ijk}=1$), which, according to the voxelization criterion, covers exactly one voxel each. For the multi-sphere representation of particles in DEM simulations, obviously, we are interested in a different optimum: The target shape should be represented with the smallest possible number of spheres such that the resulting mismatch is less than $\eta$. This condition defines a criterion for terminating the optimization algorithm.   

\section{MSS algorithm}
\subsection{General outline}
Starting point of the algorithm is the array $S$ containing the voxelized and binarized target shape according to \Cref{eq:voxel_S}. If the particle shape was obtained by X-ray Computed Tomography (CT) or Magnetic Resonance Imaging (MRI), $S$ is directly available in voxelized form. Binarization can be done by thresholding or more sophisticated approaches such as Otsu's algorithm \cite{Otsu.1979}. See \cite{Sezgin.2004} for a comprehensive overview of image thresholding techniques. In case the target shape is given in form of a polygonal surface mesh or in another representation, the voxelization requires preprocessing as described in \Cref{sec:target-particle-shape}. 

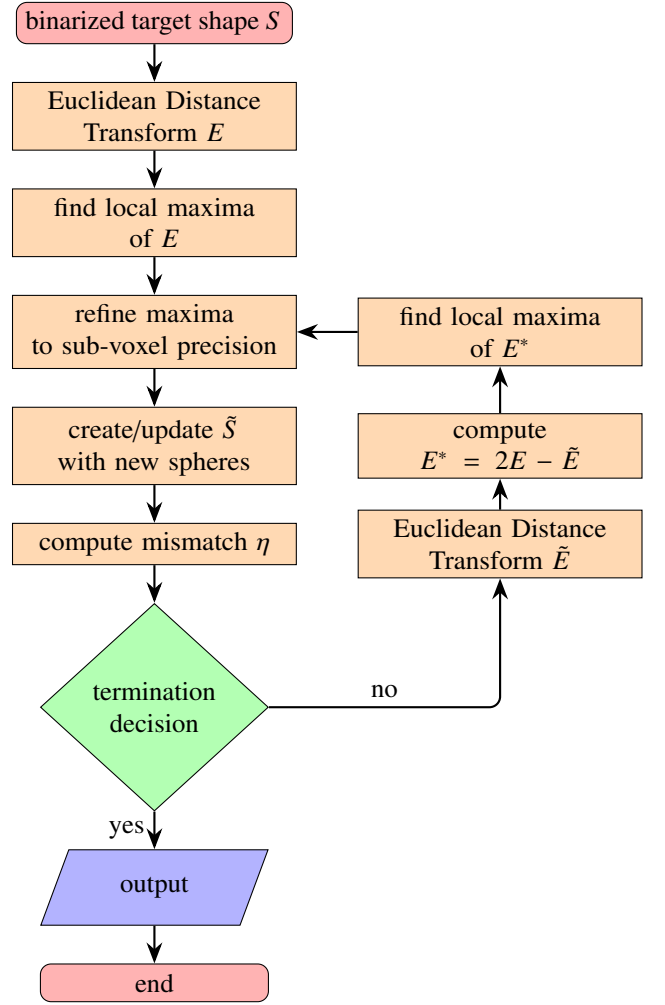
\begin{figure}[htb]
    \centering
\begin{tikzpicture}
    \node (start) [startstop] {binarized target shape $S$};
    
    \node (box1) [process,below=0.5 of start] 
    {Euclidean Distance\\ Transform $E$};
    \draw [arrow] (start) -- (box1);
    
    \node (box2) [process,below=0.5 of box1,align=center] 
    {find local maxima\\of $E$};
    \draw [arrow] (box1) -- (box2);
    
    \node (box3) [process,below=0.5 of box2,align=center] 
    {refine maxima\\to sub-voxel precision};
    \draw [arrow] (box2) -- (box3);
    
    \node (box4) [process,below=0.5 of box3,align=center,text width=3.5cm] 
    {create/update $\tilde{S}$ with new spheres};
    \draw [arrow] (box3) -- (box4);
    
    \node (box5) [process,below=0.5 of box4] 
    {compute mismatch $\eta$};
    \draw [arrow] (box4) -- (box5);
    
    \node (dec1) [decision, below=0.5 of box5, align=center] 
    {termination\\decision};
    \draw [arrow] (box5) -- (dec1);
    
    
    \node (box6) [process, right=0.8 of box5] 
    {Euclidean Distance\\Transform $\tilde{E}$};
    
    \node (box7) [process, right=0.8 of box4] 
    {compute\\$E^\ast = 2E - \tilde{E}$};
    
    \node (box8) [process, right=0.8 of box3] 
    {find local maxima\\of $E^\ast$};
    
    
    \draw [arrow] (dec1) -| node[pos=0.25,anchor=south]{no} (box6);
    \draw [arrow] (box6) -- (box7);
    \draw [arrow] (box7) -- (box8);
    \draw [arrow] (box8.west) -- (box3.east);
    
    
    \node (out1) [io, below=0.5 of dec1] {output};
    \draw [arrow] (dec1) -- node[anchor=east]{yes} (out1);
    
    \node (stop) [startstop, below=0.5 of out1] {end};
    \draw [arrow] (out1) -- (stop);
\end{tikzpicture}
\caption{Flowchart of the MSS algorithm}
\label{fig:flowchart}
\end{figure}

\Cref{fig:flowchart} shows the flowchart of MSS. First, we compute the Euclidean Distance Transform, $E$, of the target shape, $S$. This first step agrees with previously published algorithms \cite{Silin.2006, Gostick.2017, Khan.2019, Angelidakis.2021, Canbolat.2025}. Next, $E$ is used to compute an initial sphere representation $\tilde{S}$ and its Euclidean Distance Transformation $\tilde{E}$. In subsequent iterations, additional spheres are introduced to reduce the mismatch between $E$ and $\tilde{E}$. The iteration is terminated according to the criteria specified in \Cref{sec:termination_criteria}. 

\subsection{Euclidean Distance Transformation}
\label{sec:EDT}
For each element $S_{ijk}=1$ of a binary array, $S$, the Euclidean distance transformation, $E$, specifies the Euclidean distance to nearest background element $S_{ijk}=0$. The Euclidean distance transformation was originally developed in the context of digital image processing to characterize the similarity or dissimilarity of images \cite{Rosenfeld:1966}. Danielsson \cite{Danielsson:1980} introduced an efficient algorithm for approximately computing the Euclidean distance transformation in time $\mathcal{O}(N)$, where $N$ is the number of Voxels, using chamfer masks and vector propagation. This method enabled the systematic application of the Euclidean distance transformation in digital image processing. Modern algorithms \cite{Meijster:2000,Maurer.2003} compute the exact transformation while retaining linear time complexity $\mathcal{O}(N)$ and are often used in image processing libraries.

\begin{figure}[t!]
    \centering
    \mbox{\includegraphics[width=\linewidth]{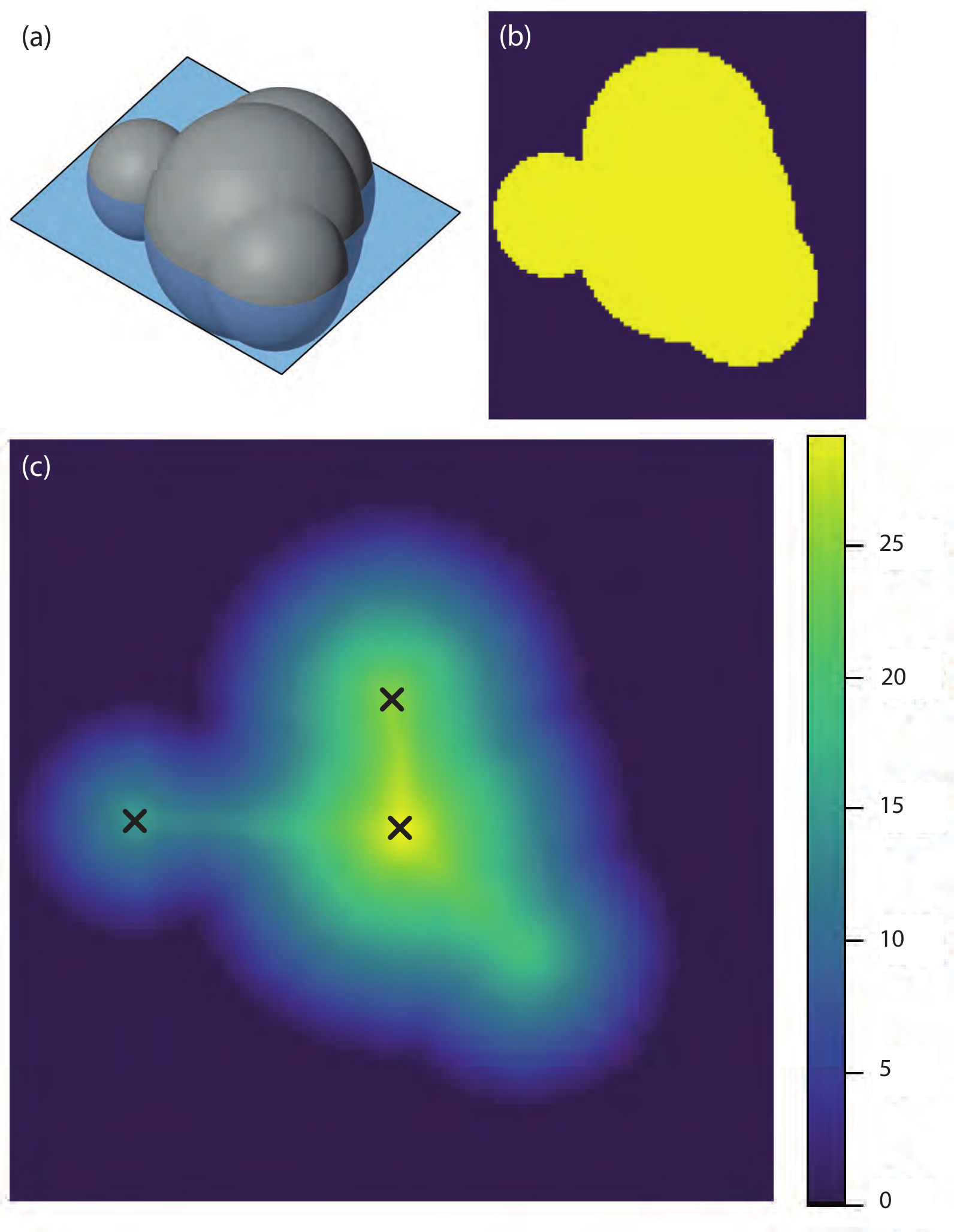}}
    \caption{3D target shape (a), a 2D voxelized and binarized array through a section along the blue plane (b), and the corresponding 2D Euclidean Distance Transform (c). The crosses indicate the local maxima of $E$ in pixel resolution. Comparison with the 3D target shape suggests that these maxima approximate the optimal location for three of the spheres to be determined.}   
    \label{fig:3D-to-2D}
    \label{fig:four_spheres_EDT}
\end{figure}

Given the voxelized target shape, $S$, as specified in \Cref{eq:voxel_S}, we obtain
\begin{equation}
\begin{split}
E_{ijk} &= \min_{(l,m,n)\in I}  \sqrt{(i-l)^2+(j-m)^2+(k-n)^2}\\
& I\equiv \left\{(l,m,n)\left| S_{lmn}=0\right.\right\}\,,
\end{split}
\label{eq:EDT-def}
\end{equation}
where the index set $I$ contains all background voxels. For illustration, \Cref{fig:3D-to-2D} shows a target shape, ${\cal S}$, consisting of four mutually penetrating spheres, together with the voxelized and binarized array $S$ and the corresponding $E$.

\Cref{fig:3D-to-2D} illustrates the Euclidean distance transformation in two dimensions for ease of visualization, however, MSS operates on three-dimensional target shapes. The subsequent steps of the algorithm are illustrated using  the same example.

\subsection{Geometric morphology of the Euclidean Distance Transform at voxel resolution and sub-voxel resolution}
\label{sec:algo:locmax}

The Euclidean Distance Transform assigns each voxel its distance to the background. A voxel $S_{ijk}$ is an element of the three-dimensional array $S$ identified by its indices $(i,j,k)$. By construction (see \Cref{sec:voxel}), these indices are associated with the centers of the corresponding voxels. They can, therefore, serve simultaneously as array indices and as coordinates on a dimensionless voxel lattice. 

Since $(i,j,k)$ are integers, at this level (the voxel-resolution level), the computation of the locations of the maxima of $E$ cannot be more accurate than the voxel size. Local maxima of $E$ at voxel resolution are, thus, those voxels $i^\voxmax,j^\voxmax, k^\voxmax$ whose $3\times 3\times 3$ neighborhood does not contain elements larger than $E_{i^\voxmax,j^\voxmax, k^\voxmax}$. Only inner voxels of $E$ can contain local maxima, as its outer elements are of value 0, by definition, see \Cref{eq:S-rand}. Note that this definition includes plateaus where adjacent voxels have equal $E$. \Cref{fig:3D-to-2D}c shows $E$ with the locations of the local maxima marked by crosses.

According to its definition, in close vicinity of a local maximum, $E$ decreases linearly with increasing distance to the maximum:
\begin{equation}
    E_{ijk}=a+b\left|\begin{pmatrix}i\\j\\k\end{pmatrix}-\begin{pmatrix}i^\truemax\\j^\truemax\\k^\truemax\end{pmatrix}\right| \,,
\end{equation}
Where, in general, the slope $b$ is anisotropic,
\begin{equation}
    b=b\left(\frac{\begin{pmatrix}i\\j\\k\end{pmatrix}-\begin{pmatrix}i^\truemax\\j^\truemax\\k^\truemax\end{pmatrix}}{\left|\begin{pmatrix}i\\j\\k\end{pmatrix}-\begin{pmatrix}i^\truemax\\j^\truemax\\k^\truemax\end{pmatrix}\right|}\right)\,.
\end{equation}
The sub-voxel maximum $(i^\truemax,j^\truemax,k^\truemax)$ is expressed in the same voxel-aligned coordinate system but may assume non-integer values. To approximate $E_{i^\truemax j^\truemax k^\truemax}$, we apply the following approach: For each voxel $\left(i^\voxmax,j^\voxmax, k^\voxmax\right)$ corresponding to a local maximum of $E$ at \emph{voxel resolution}, we analyze the three coordinate directions $i$, $j$, and $k$ separately. Consider first the coordinate $i$ and the five points (\Cref{{fig:sketch-lin-fit}}a)
\begin{figure}[htb!]
    \centering\includegraphics[width=1.0\columnwidth]{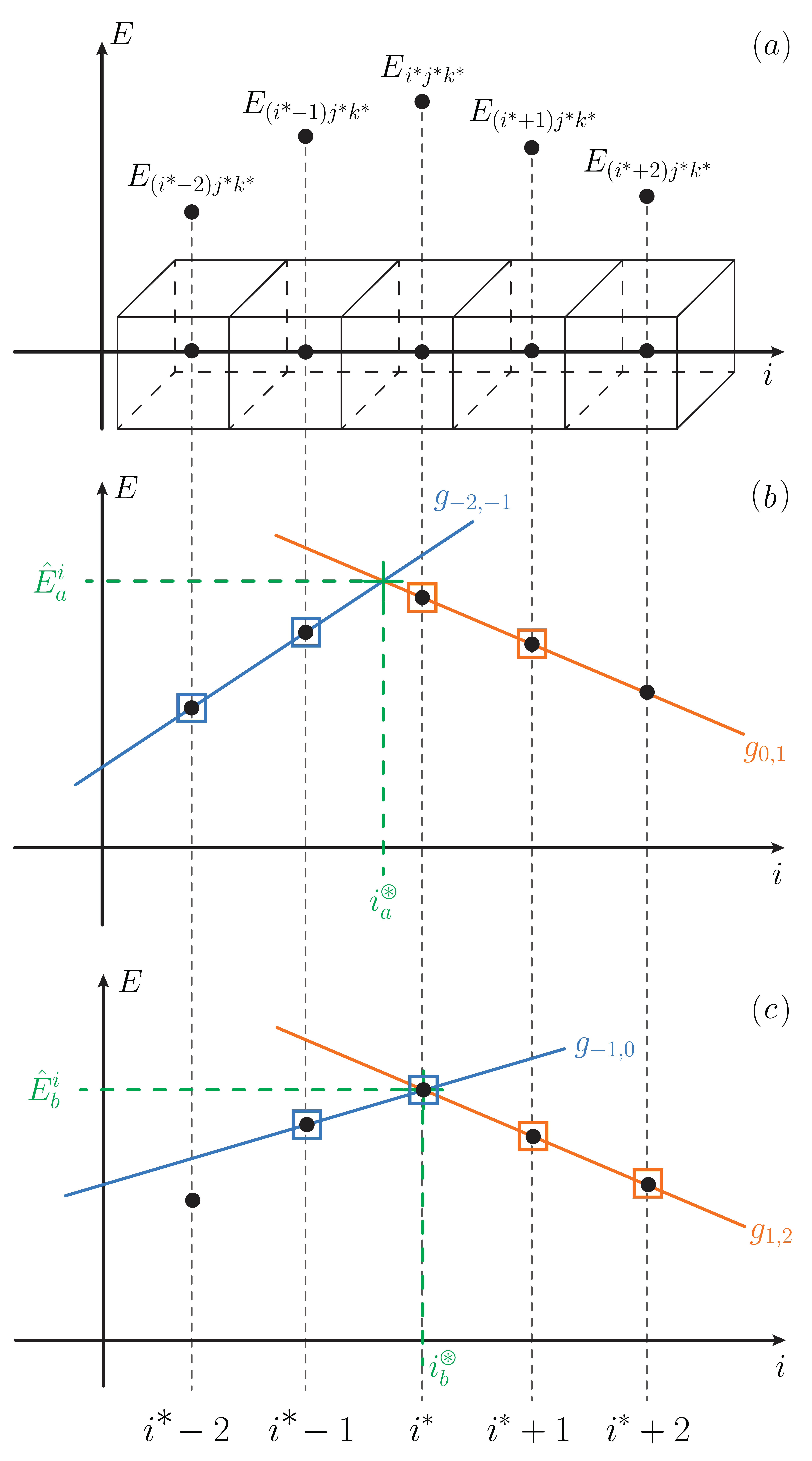}

    \caption{Computation of the location $i^\truemax$ of a local maximum of E at sub-voxel resolution. (a) specifies the considered data points. (b) and (c) illustrate the pairs of straight lines $(g_{-2,-1};g_{0,1})$ and $(g_{-1,0};g_{1,2})$, defining $i^\truemax_a$ and $i^\truemax_b$ respectively.
    \label{fig:sketch-lin-fit}}
\end{figure}
\begin{equation}
    E_{i,j^\voxmax, k^\voxmax} \quad \text{with}\quad i^\voxmax-2 \le i \le i^\voxmax+2\,.
    \label{eq:5points}
\end{equation}
The straight line $g_{-2,-1}$ (\Cref{{fig:sketch-lin-fit}}b) is defined by the points $\left(i^\voxmax-2, E_{i^\voxmax-2,j^\voxmax,k^\voxmax}\right)$ and $\left(i^\voxmax-1, E_{i^\voxmax-1,j^\voxmax,k^\voxmax}\right)$. Another straight line $g_{0,1}$ is defined by $\left(i^\voxmax, E_{i^\voxmax,j^\voxmax,k^\voxmax}\right)$ and $\left(i^\voxmax+1,E_{i^\voxmax+1,j^\voxmax,k^\voxmax}\right)$.
If these lines are not parallel, they intersect at 
\begin{equation}
    i^\truemax_a=i^\voxmax+\lambda_a
\end{equation}
with
\begin{equation}
    \lambda_a \equiv\frac{\left(E_{i^\voxmax,j^\voxmax,k^\voxmax}-E_{i^\voxmax-1,j^\voxmax,k^\voxmax}\right)-\left(E_{i^\voxmax-1,j^\voxmax,k^\voxmax}-E_{i^\voxmax-2,j^\voxmax,k^\voxmax}\right)}{\left(E_{i^\voxmax-1,j^\voxmax,k^\voxmax}-E_{i^\voxmax-2,j^\voxmax,k^\voxmax}\right)-\left(E_{i^\voxmax+1,j^\voxmax,k^\voxmax}-E_{i^\voxmax,j^\voxmax,k^\voxmax}\right)}\,.
\end{equation}
The corresponding $E$-value at the spatial location $i^\truemax_a$ is 
\begin{multline}
    \hat{E}^i_a=E_{i^\voxmax,j^\voxmax,k^\voxmax}+\lambda_a\left(E_{i^\voxmax-1,j^\voxmax,k^\voxmax}-E_{i^\voxmax-2,j^\voxmax,k^\voxmax}\right)+\\
    \left[\left(E_{i^\voxmax-1,j^\voxmax,k^\voxmax}-E_{i^\voxmax-2,j^\voxmax,k^\voxmax}\right)-\left(E_{i^\voxmax,j^\voxmax,k^\voxmax}-E_{i^\voxmax-1,j^\voxmax,k^\voxmax}\right)\right]
\end{multline}
Note that the superscript $i$ is not an index, but it stands for the direction $i$ in the voxel-centered coordinate system. Analogously, we define $\hat{E}^j_a$ and $\hat{E}^k_a$ for the $j$ and $k$ directions.

If the lines are parallel, our local maximum is part of a plateau where adjacent voxels have identical values. In this case, the maximum at sub-voxel resolution agrees with the maximum at voxel resolution and, thus,
\begin{align}
    i^\truemax_a &=i^\voxmax\\
    \hat{E}^i_a &= E_{i^\voxmax,j^\voxmax,k^\voxmax}\,.
\end{align}

Alternatively, we can define another pair of intersecting lines from the points selected in \Cref{eq:5points}, see \Cref{{fig:sketch-lin-fit}}c. Namely, $g_{-1,0}$, specified by $\left(i^\voxmax-1, E_{i^\voxmax-1,j^\voxmax,k^\voxmax}\right)$ and $\left(i^\voxmax, E_{i^\voxmax,j^\voxmax,k^\voxmax}\right)$, and $g_{1,2}$ specified by $\left(i^\voxmax+1, E_{i^\voxmax+1,j^\voxmax,k^\voxmax}\right)$ and $\left(i^\voxmax+2, E_{i^\voxmax+2,j^\voxmax,k^\voxmax}\right)$. This pair of lines intersects at
\begin{equation}
    i^{\truemax}_b=i^\voxmax+\lambda_b
\end{equation}
with
\begin{equation}
    \lambda_b\equiv\frac{\left(E_{i^\voxmax+1,j^\voxmax,k^\voxmax}-E_{i^\voxmax,j^\voxmax,k^\voxmax}\right)-\left(E_{i^\voxmax+2,j^\voxmax,k^\voxmax}-E_{i^\voxmax+1,j^\voxmax,k^\voxmax}\right)}{\left(E_{i^\voxmax,j^\voxmax,k^\voxmax}-E_{i^\voxmax-1,j^\voxmax,k^\voxmax}\right)-\left(E_{i^\voxmax+2,j^\voxmax,k^\voxmax}-E_{i^\voxmax+1,j^\voxmax,k^\voxmax}\right)}\,.
\end{equation}
The corresponding $E$-value reads
\begin{equation}
    \hat{E}_b^i=E_{i^\voxmax,j^\voxmax,k^\voxmax}+\lambda_b\left(E_{i^\voxmax,j^\voxmax,k^\voxmax}-E_{i^\voxmax-1,j^\voxmax,k^\voxmax}\right)\,,
\end{equation}
provided, the lines are not parallel. Otherwise
\begin{align}
    i^\truemax_b &=i^\voxmax\\
    \hat{E}^i_b &= E_{i^\voxmax,j^\voxmax,k^\voxmax}\,.
\end{align}

By construction, either $\hat{E}^i_a = E_{i^\voxmax,j^\voxmax,k^\voxmax}$ or $\hat{E}^i_b = E_{i^\voxmax,j^\voxmax,k^\voxmax}$, irrespective whether any pair of lines intersect. On the other hand, the maximum at sub-voxel resolution cannot be smaller than the maximum at voxel resolution. Therefore, we obtain at sub-voxel resolution
\begin{equation}
    \left(i^\truemax, \hat{E}^i\right) =
    \begin{cases}
        \left(i^\truemax_a, \hat{E}^i_a\right) & \text{if}~~ \hat{E}^i_a > E_{i^\voxmax,j^\voxmax,k^\voxmax}\\[0.1cm]
        \left(i^\truemax_b, \hat{E}^i_b\right) & \text{otherwise}\,.
    \end{cases}
\end{equation}

\begin{figure}[htb]
    \centering
    \includegraphics[width=0.95\columnwidth,angle=0]{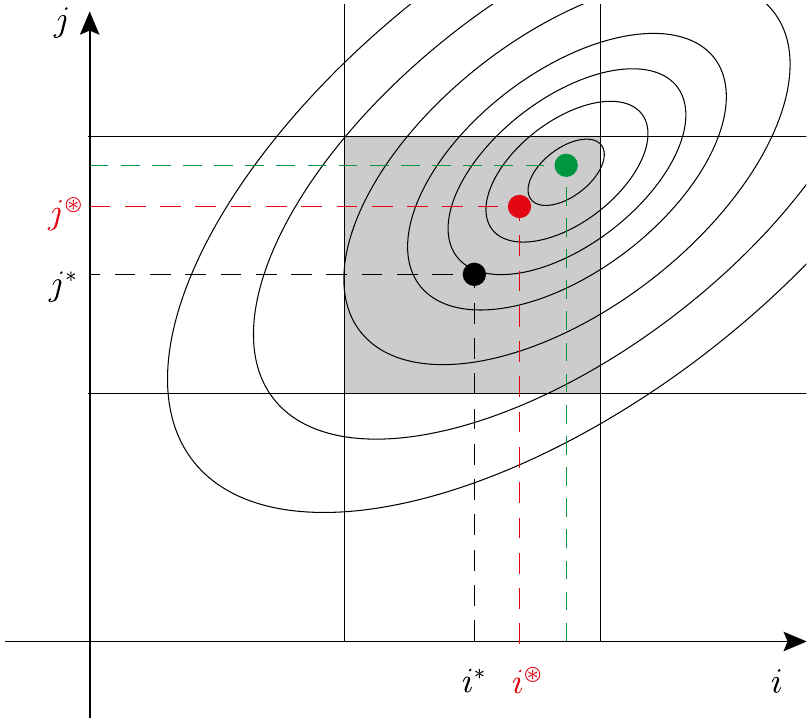}
    \caption{Section of $E$ along the $i-j$ plane. Black curves show iso-lines of constant E. Black dot: center of the grey shaded voxel $i^\voxmax,j^\voxmax,k^\voxmax$ (location of a local $E$ maximum in voxel resolution). Red dot: $\hat{E}$ maximum in sub-voxel resolution $(i^\truemax,j^\truemax,k^\truemax)$; green dot: true yet unknown location of the $E$ maximum.
    \label{fig:sketch-plane-iter}}
\end{figure}

The process described so far delivers the location $i^\truemax$ of the $E$-maximum with respect to the $i$ coordinate, with the constraint that the other coordinates assume the values $j^\voxmax$ and $k^\voxmax$. The same can be done for the coordinates $j$ and $k$, to obtain $j^\truemax$ and $k^\truemax$ and the corresponding $\hat{E}^j$ and $\hat{E}^k$. In view of the aforementioned constraint, the vector $(i^\truemax,j^\truemax,k^\truemax)$  cannot maximize $E$ but only approximate the maximum. This is also clear from the fact that, in general, $E_i^\truemax$, $E_j^\truemax$, and $E_k^\truemax$ differ from one another. \Cref{fig:sketch-plane-iter} sketches the location of a local $E$-maximum in the $i-j$ plane in voxel resolution $\left(i^\voxmax, j^\voxmax, k^\voxmax\right)$ and sub-voxel resolution $\left(i^\truemax, j^\truemax, k^\truemax\right)$, together with the true (yet unknown) maximum. 

The position of the $E$-maximum in sub-voxel resolution in physical space is obtained by multiplying the non-integer index vector by the voxel size,
\begin{equation}
    \label{eq:truemax-r}
\vec{r}^{\,\truemax}=\latconst
\begin{pmatrix}
i^\truemax\\ j^\truemax\\ k^\truemax
\end{pmatrix}\,.
\end{equation}
The maximum is approximated by
\begin{equation}
    \label{eq:truemax-E}
E^\truemax =\latconst \max\left\{E_i^\truemax,E_j^\truemax,E_k^\truemax\right\}\,.
\end{equation}
This choice is reasonable since we know that $E^\truemax$ is achieved by one of the vectors $\left(i^\truemax,j^\voxmax,k^\voxmax\right)$ or $\left(i^\voxmax,j^\truemax,k^\voxmax\right)$ or $\left(i^\voxmax,j^\voxmax,k^\truemax\right)$, therefore, the true maximum cannot be smaller.

\begin{figure}[htb]
    \centering
    \includegraphics[width=\linewidth]{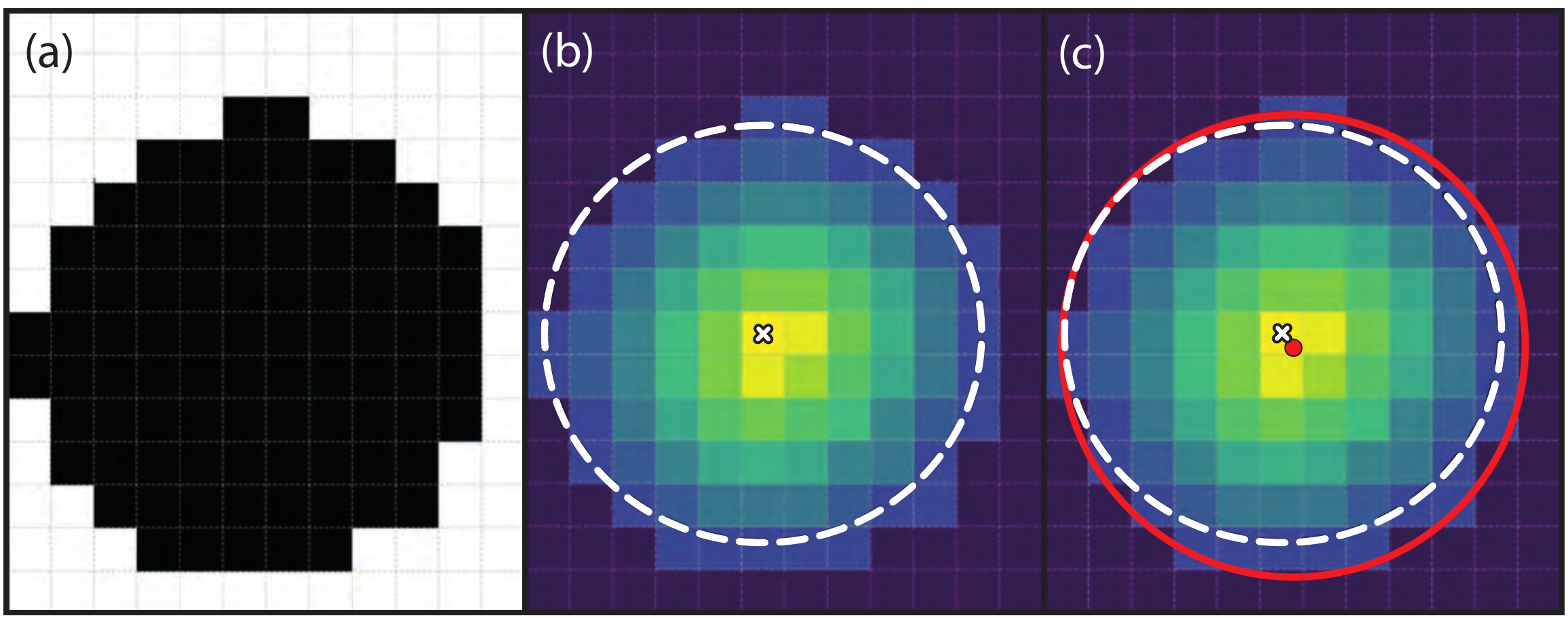}
    \caption{
    Sub-voxel precision versus voxel precision. (a) voxelized and binarized 2D section through a target shape. (b) Location of the maximum of $E$ in voxel precision, $\left(\latconst i^\voxmax, \latconst j^\voxmax, \latconst k^\voxmax\right)$ and a circle of radius $\latconst E_{i^\voxmax j^\voxmax k^\voxmax}$. (c) Location of the maximum of $E$ in sub-voxel precision,  $\vec{r}^{\,\truemax}$, and the red circle of radius $E^\truemax$}
    \label{fig:sketch-fit}
\end{figure}

\Cref{fig:sketch-fit}
shows a 2d section trough a voxelized target shape and the locations of the maximum of $E$ in voxel precision, $\left(\latconst i^\voxmax, \latconst j^\voxmax, \latconst k^\voxmax\right)$ and in sub-voxel precision, $\vec{r}^{\,\truemax}$. The white circle has the radius $\latconst E_{i^\voxmax j^\voxmax k^\voxmax}$ (voxel precision), the red circle has the radius $\latconst E^\truemax$ (sub-voxel precision).

The described approximation can be further refined: To this end, we shift the grid origin to $\vec{r}^{\,\truemax}$ (the red dot in \Cref{fig:sketch-plane-iter}) and repeat the procedure. This iterative process progressively approximates the location of the optimum. Each iteration requires repeated computation of the Euclidean Distance Transform which is the most computationally expensive step. Moreover, when the grid is shifted, the Euclidean Distance Transform must be computed separately for each local maximum of $E$ (at voxel resolution). Consequently, each local maximum is associated with its own grid.

In all cases studied, it turned out to be more efficient to reduce the voxel size $\latconst$ instead of iterating the computation in sub-voxel precision.

\subsection{Sphere representation of the target shape}
\label{sec:sphere_rep}
\begin{figure}[h]
    \centering
    \includegraphics[width=\linewidth]{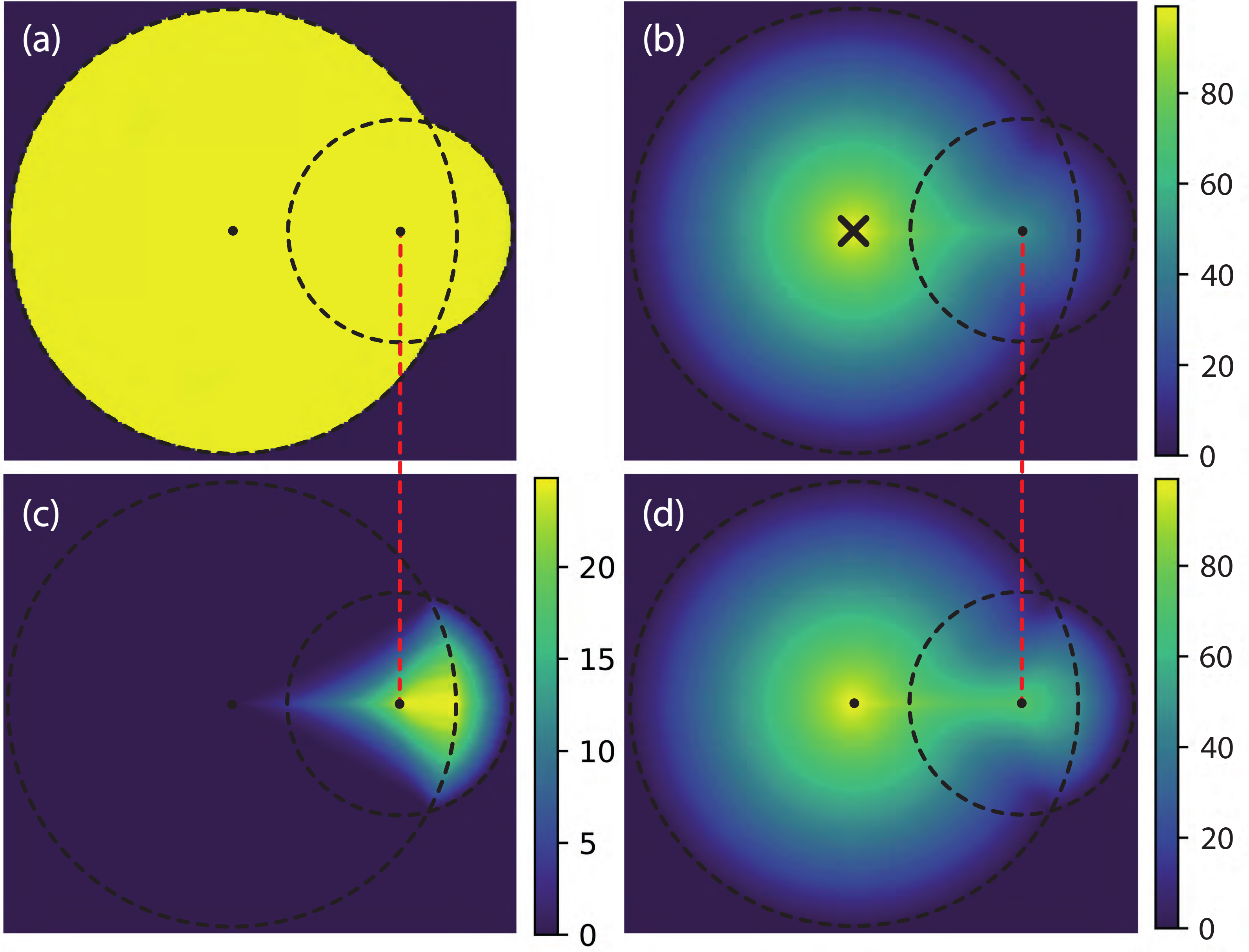}
    \caption{Geometric interpretation of $E^*$. (a) The target $\cal{S}$ consists of two intersecting spheres. (b) Its $E$ reveals one local maximum, indicated by a black cross. Thus, $\tilde{\cal S}$ consists of a single sphere. (c) The difference $E-\tilde{E}$ indicates that $\cal{S}$ can be improved by adding a sphere, however, the location of the maximum (red line) does not agree with the maximum of $E-\tilde{E}$. (d) $2E-\tilde{E}$ reveals local maxima at the centers of both spheres of $\cal{S}$. 
    }
    \label{fig:algo:fedt:expl}
\end{figure}

At each position $\vec{r}^{\,\truemax}$ corresponding to a local maximum of $E$ (\Cref{eq:truemax-r}), we place a sphere with radius $E^\truemax$ given by \Cref{eq:truemax-E}. The union of these spheres, $\tilde{\cal{S}}^0$, defined by \Cref{eq:sphereRep} is the initial multi-sphere representation of $\cal{S}$. Starting from $\tilde{\cal{S}}^0$, spheres are added iteratively to improve $\tilde{\cal{S}}$ until one of the termination criteria defined in \Cref{sec:termination_criteria} is satisfied. Each iteration comprises the following steps:
\begin{enumerate}[leftmargin=1.7em]
    \item[$i)$] Compute the voxelized form $\tilde{S}$ of $\tilde{\cal{S}}$ according to \Cref{eq:voxel_S}. 
    \item [$ii)$] Compute the Euclidean Distance Transformation $\tilde{E}$ according to \Cref{eq:EDT-def}.
    \item [$iii)$] Compute 
    \begin{equation}
        E^\ast \equiv E + \left(E -\tilde{E}\right) = 2E -\tilde{E}
        \label{eq:enhanced_field}
    \end{equation}
    and the local maxima of $E^*$ in sub-voxel resolution. \Cref{fig:algo:fedt:expl} illustrates the geometric interpretation of $E^\ast$.
    \item [$iv)$] Add spheres to $\tilde{\cal{S}}$, whose centers are located at the local maxima of $E^\ast$ at sub-voxel precision and whose radii are obtained from $E$, evaluated at these center locations using the same linear interpolation as described in \Cref{sec:algo:locmax}.
\end{enumerate}

\begin{figure}[h]
    \centering
    \includegraphics[width=\linewidth]{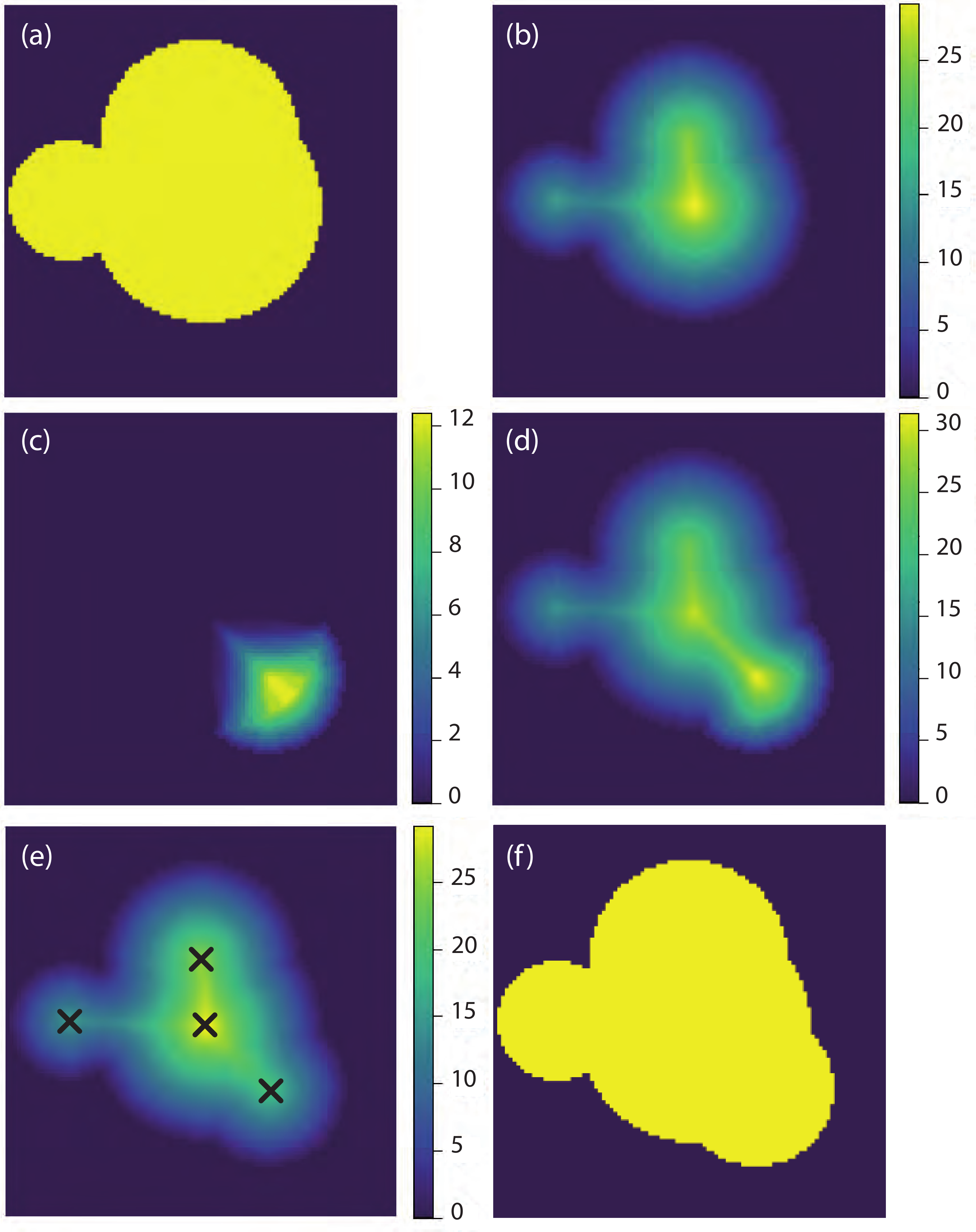}
    \caption{(Illustration of the first iteration. a) Initial multi-sphere representation $\tilde{\cal{S}}^0$ for $\cal{S}$ shown in \Cref{fig:four_spheres_EDT}a. Comparison with \Cref{fig:four_spheres_EDT}b reveals that one of the spheres is not captured by $\tilde{\cal{S}}^0$. (b) Euclidean Distance Transformation $\tilde{E}$ computed from $\tilde{\cal{S}}^0$. (c) Difference between $E$ (see \Cref{fig:four_spheres_EDT}c) and $\tilde{E}$. (d) The field $2E-\tilde{E}$ differs from $E$ (panel c) in one additional local maximum. (e) The local maxima indicate the locations of spheres added to $\tilde{\cal{S}}$. (f) The resulting $\tilde{\cal{S}}$ agrees well with $\cal{S}$ shown in \Cref{fig:four_spheres_EDT}(b).
} 
    \label{fig:SRM-EDT}
\end{figure}

\noindent \Cref{fig:SRM-EDT} demonstrates the first iteration, steps $i$ to $iv$.

Note that in step $iv$, the locations of spheres are the local maxima of $E^\ast$ (at sub-voxel resolution). Their radii, however, must be determined from $E$ through linear interpolation, since $E^*$ does not represent an Euclidean Distance Transform and, therefore, does not provide geometrically meaningful sphere radii.

In some cases, the positions of the new spheres can be very close to spheres that are already part of $\tilde{\cal{S}}$. Adding such spheres would increase the number of spheres in $\tilde{\cal{S}}$ while only marginally reducing the mismatch $\eta$. This can be avoided by requiring a minimal distance, $\delta_{\min}$, between any pair of spheres in $\tilde{\cal{S}}$. Thus, local maxima of $E^\ast$ which are closer than $\delta_\text{min}$ to any sphere in $\tilde{\cal{S}}$ are disregarded.

\subsection{Termination criteria and geometric constraints}
\label{sec:termination_criteria}

The iterative refinement of the multi-sphere representation is terminated once predefined quality or complexity criteria are satisfied. A primary criterion is the mismatch between the target shape, $S$, and the reconstruction, $\tilde{S}$, quantified by the measures introduced in \Cref{sec:mismatch}. The iteration terminates when the mismatch $\eta$ drops below a prescribed threshold. Other reasonable termination criteria involve the total number of spheres in $\tilde{\cal{S}}$ and their minimum radius. All these criteria can be applied individually or combined.

An important application of MSS is DEM simulations of particles that undergo plastic deformations, including particle fragmentation. In such cases, geometric constraints must be imposed: the spheres of a multi-sphere particle model must not penetrate specified planes. Such planes can be the surface mesh, system walls, or the fragmentation plane.

The need for this constraint follows from the contact mechanics. The elastic restoring force between identical\footnote{The restriction to identical spheres was introduced solely to keep the numerical example simple.} spheres in contact is described by Hertz's contact force \cite{hertz:1882},
\begin{equation}
F^\text{Hertz}=\frac{2Y}{3\left(1-\nu^2\right)}\sqrt{\frac{R}{2}}\xi^\frac{3}{2}\,,
    \label{eq:Hertz}
\end{equation}
where $\xi\equiv 2R-\left|\vec{r}_1-\vec{r}_2\right|$ characterizes the deformation of the spheres located at $\vec{r}_1$ and $\vec{r}_2$. Typical values of the Young's modulus and the Poisson's ratio are $Y=(10\dots200)\times 10^9\, \text{Pa}$ and $\nu=0.1\dots 0.4$. In DEM simulations, even small deformations can lead to significant forces. For instance, for $R=1\,\text{cm}$, $Y=100\,\text{GPa}$, and $\nu=0.3$, a deformation as small as $\xi=10\,\mu\text{m}$ results in $F^\text{Hertz}\approx 164\,\text{N}$. Therefore, even a slightly incorrect compression of particles due to shape changes of particles which are in contact or in close proximity can easily destabilize a DEM simulation, and must, thus, be avoided. 

The MSS algorithm described so far does not consider this constraint, as discretization errors can produce spheres that extend beyond the target shape by a maximum of one grid length $\latconst$. If this is not acceptable --as in the case of particle fragmentation--, the following modification can be applied: For each sphere of the multi-sphere representation we compute the distance $\tilde{R}$ from the center of the sphere to the nearest vertex or face of the mesh that defines the target shape. If this distance is smaller than the radius of the sphere $R$, we set the radius of the sphere to $R\xrightarrow{}\tilde{R}$. 

\subsection{Implementation of MSS}
MSS is implemented in \texttt{Python} and in \texttt{C++}. The algorithm can be integrated into DEM codes such as MercuryDPM, enabling the use of MSS directly within DEM simulations. Implementation details are given in \cite{Buchele.2026, Moradian.2026}.

\section{Performance\label{sec:performance}}

\subsection{Benchmarking}

We compare the performance of MSS against other multi-sphere generators reported in the literature \cite{Favier:1999, Ferellec.2010, Angelidakis.2021}. In practice, however, only the \textsc{Clump} algorithm \cite{Angelidakis.2021} is used as a reference in this study. \textsc{Clump} is a current, widely used open-source program that has been extensively and systematically benchmarked against numerous alternative methods \cite{Angelidakis.2021} where it was identified as the best-performing algorithm available at the time. Nevertheless, we demonstrate that MSS outperforms \textsc{Clump} with respect to both the quality of the generated multisphere representations and computational efficiency.

We use the Dice-S{\o}rensen coefficient introduced in \cite{Dice.1945} (as cited in \cite{Levy.2025}),
\begin{equation}
    D \equiv 1 -
    \frac{2 \sum\limits_{i,j,k} \left( S_{ijk} \cdot \tilde{S}_{ijk} \right) }{\sum\limits_{i,j,k} S_{ijk} + \sum\limits_{i,j,k} \tilde{S}_{ijk} },
    \label{eq:mismatch-dice}
\end{equation}
to quantify the scale-independent mismatch between the multi-sphere representation $\tilde{\cal{S}}$ of a given particle and the particle $\cal{S}$ itself. 

While MSS does not require any tuning parameters beyond the termination criteria defined in \Cref{sec:termination_criteria}, the reference algorithm \textsc{Clump} relies on several user-defined parameters that influence the reconstruction quality. To ensure a fair comparison, we perform a parameter sweep and select the configuration yielding the best reconstruction accuracy for each test case. The optimal parameter values used for comparison are shown in \Cref{app:parameters}. The absence of adjustable parameters is a key advantage of MSS, since their optimal values are generally unknown and must be identified empirically for each application through trial-and-error procedures.

\subsection{Simple object: ellipsoid}

\begin{figure}[htbp]
    \centering
    \includegraphics[width=\linewidth]{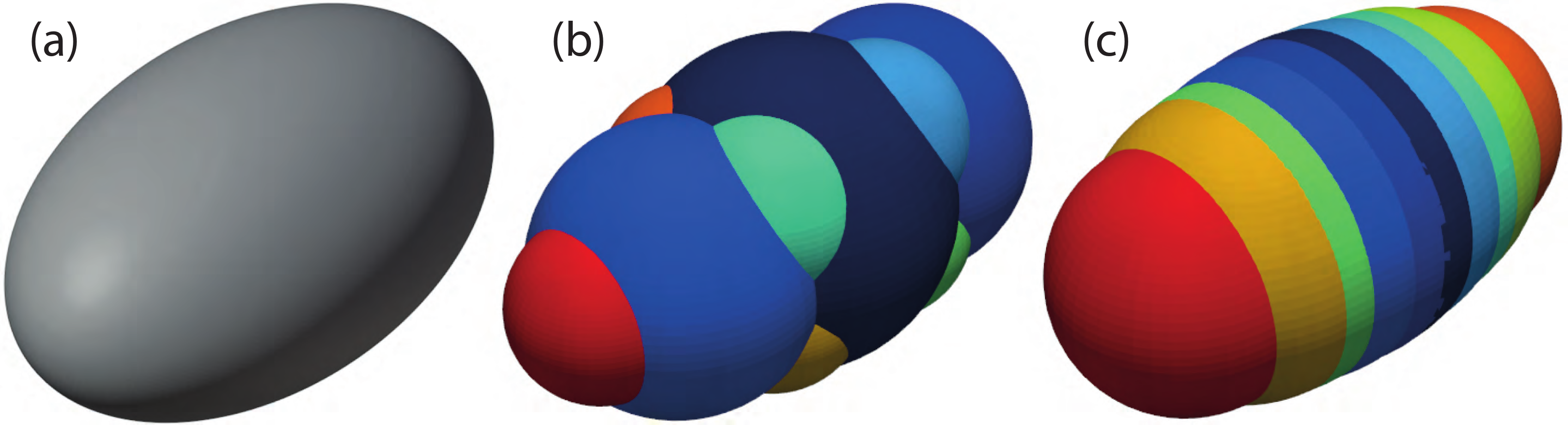}
\vspace*{0.0cm}

{\footnotesize 
\begin{tabular}{|l|l|l|}
\hline
& reference & MSS \\
\hline
mismatch $D$ & 0.10 & 0.0626 \\
runtime & 2.71\,sec & 1.3\,sec\\
\hline
\end{tabular}}
    \caption{Benchmark ellipsoid. a) target shape $\cal{S}$; b) multi-sphere representation obtained by the reference algorithm; c) multi-sphere representation $\tilde{\cal{S}}$  generated by MSS. The benchmark results for the relative mismatch, $D$, and the runtime are also given.}
    \label{fig:ellipsoid}
\end{figure}

We consider an ellipsoid as target shape $\cal{S}$, which is voxelized into a $[209 \times 107 \times 107]$ voxel array $S$. This benchmark was introduced in \cite{Angelidakis.2021}. The task is to represent $\cal{S}$ as accurately as possible by means of 10 spheres. \Cref{fig:ellipsoid}
shows the target shape together with the multi-sphere representations generated by the reference algorithm and MSS, respectively. The benchmark results are also shown (runtime excluding visualization). 

\Cref{fig:ellipsoid} reveals two significant differences between MSS and the reference algorithm: Unlike the reference, MSS preserves the symmetry of the target shape. Since $\cal{S}$ is cylindrically symmetric, the produced multi-sphere representation $\tilde{\cal{S}}$ is likewise cylindrically symmetric. Also in contrast to the reference, MSS avoids spurious small spheres that increase the roughness of the object and thus the mismatch. Even if the extra mismatch is not too large, the surface texture due to small spheres can result in undesired behavior in DEM simulations \cite{Feng.2023, Zhao.2023}.

\subsection{Complex object: human femur bone}

\begin{figure}[htb]
    \centering
        \includegraphics[width=\linewidth]{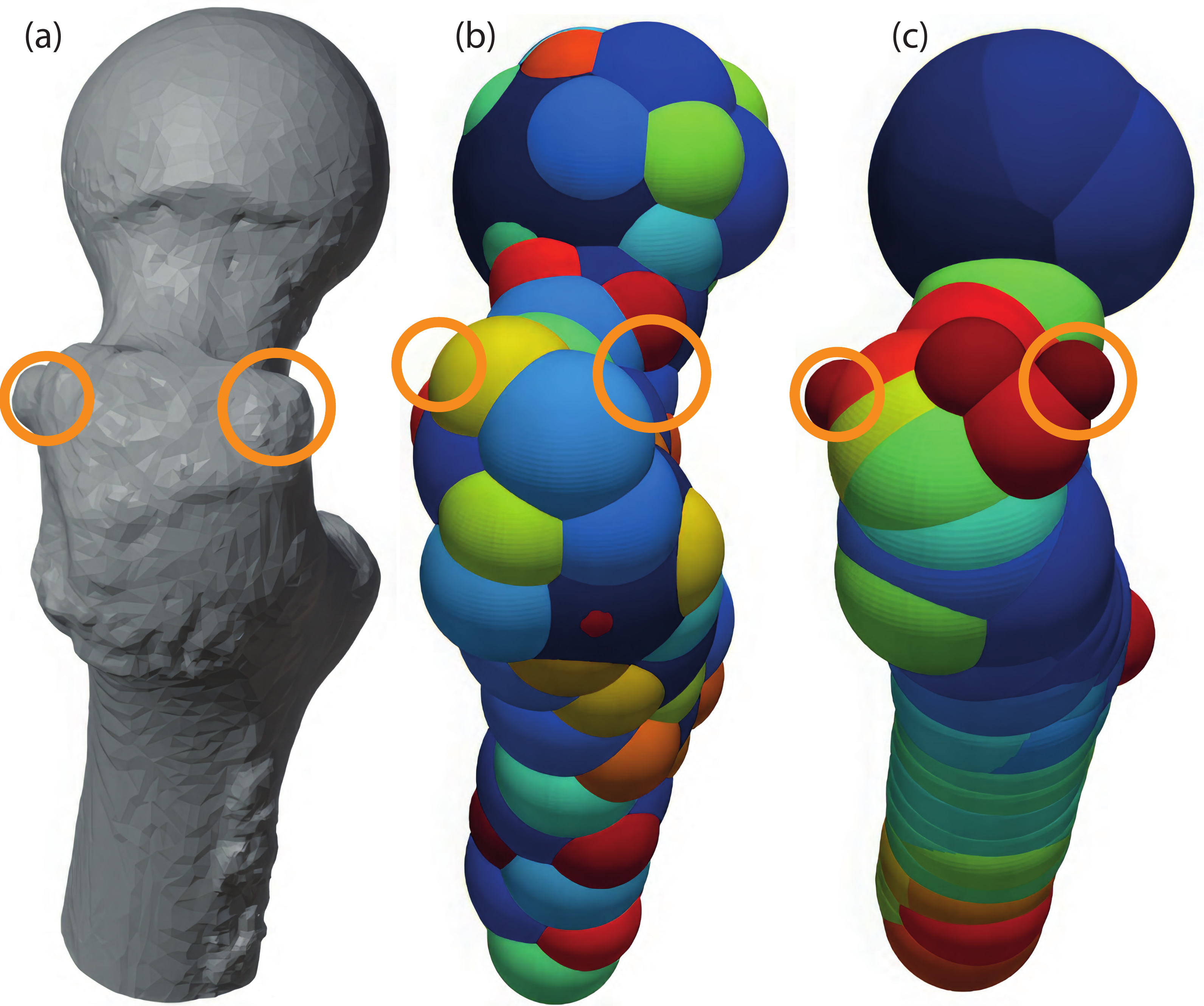}

        {\footnotesize \begin{center}
\begin{tabular}{|l|l|l|}
\hline
& reference & MSS \\
\hline
mismatch $D$ & 0.2138 & 0.1760 \\
runtime & 21.58\,sec & 1.88\,sec\\
\hline
\end{tabular}
\end{center}
}
    \caption{Benchmark human femur bone. a) target shape $\cal{S}$; multi-sphere representations, $\tilde{\cal{S}}$, using 70 spheres by the reference algorithm (b) and by MSS (c). The multi sphere representations reveal different features discussed in the text. Orange circles highlight the lesser greater trochanters.}
    \label{fig:femur}
\end{figure}

The human femur bone shown in \Cref{fig:femur} 
is represented by a multisphere model consisting of 70 spheres. This problem was previously proposed as a benchmark test in Ref. \cite{Angelidakis.2021}. The input geometry is a 3D model reconstructed from X-ray CT data and voxelized into a $[127 \times 107 \times 205]$ array $S$. Once again, MSS achieves a smaller mismatch at significantly reduced computational cost. \Cref{fig:femur} reveals several interesting differences. The reference algorithm generates spurious spheres that lead to an unrealistic surface texture. This effect is particularly pronounced in the upper part (the femoral head), which appears nearly spherical in both the CT data and the MSS reconstruction, but exhibits a bumpy surface in the reference reconstruction. A similar observation applies to the lower part (the femoral shaft), which is nearly cylindrical in the CT data and the MSS reconstruction, but not in the reference reconstruction. Functionally important anatomical features, such as the lesser and greater trochanters (highlighted by orange circles), are clearly resolved in the CT data and the MSS reconstruction but are largely suppressed in the reference reconstruction.

\subsection{Systematic studies: Sand atlas}

\begin{figure}[htbp]
    \centering
        \includegraphics[width=\linewidth]{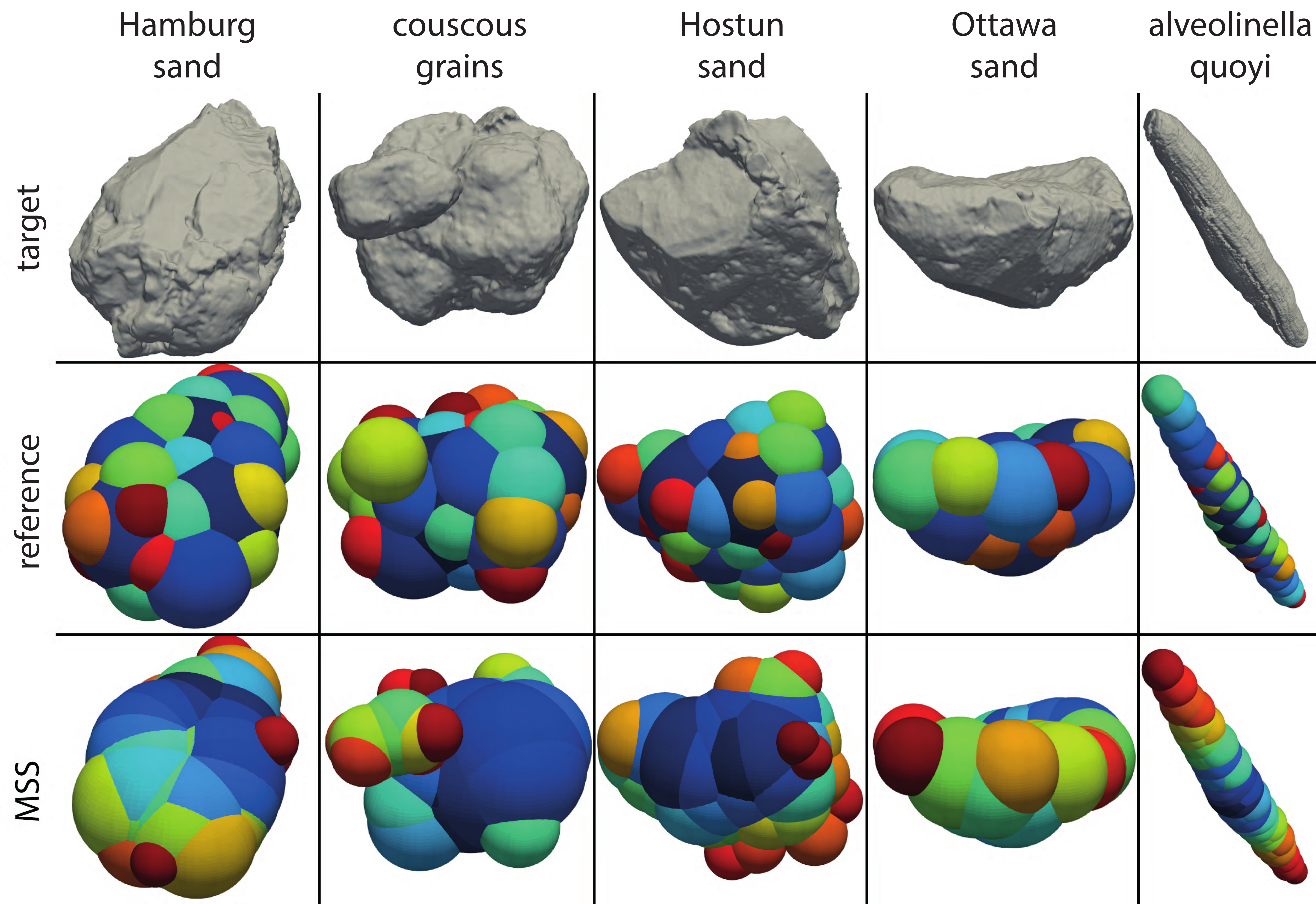}
    \caption{Exemplary particle shapes from the Sand Atlas \cite{Vego.2025}. From left to right: Hamburg sand \cite{Milatz2021}, couscous grains \cite{Vego.2023}, Hostun sand \cite{Wiebicke.2017}, Ottawa sand \cite{Saadatfar.2012}, alveolinella quoyi grains \cite{Luijmes.2024}.}
    \label{fig:sand-examples}
\end{figure}

\begin{figure}[htbp]
    \centering
        \includegraphics[width=\linewidth]{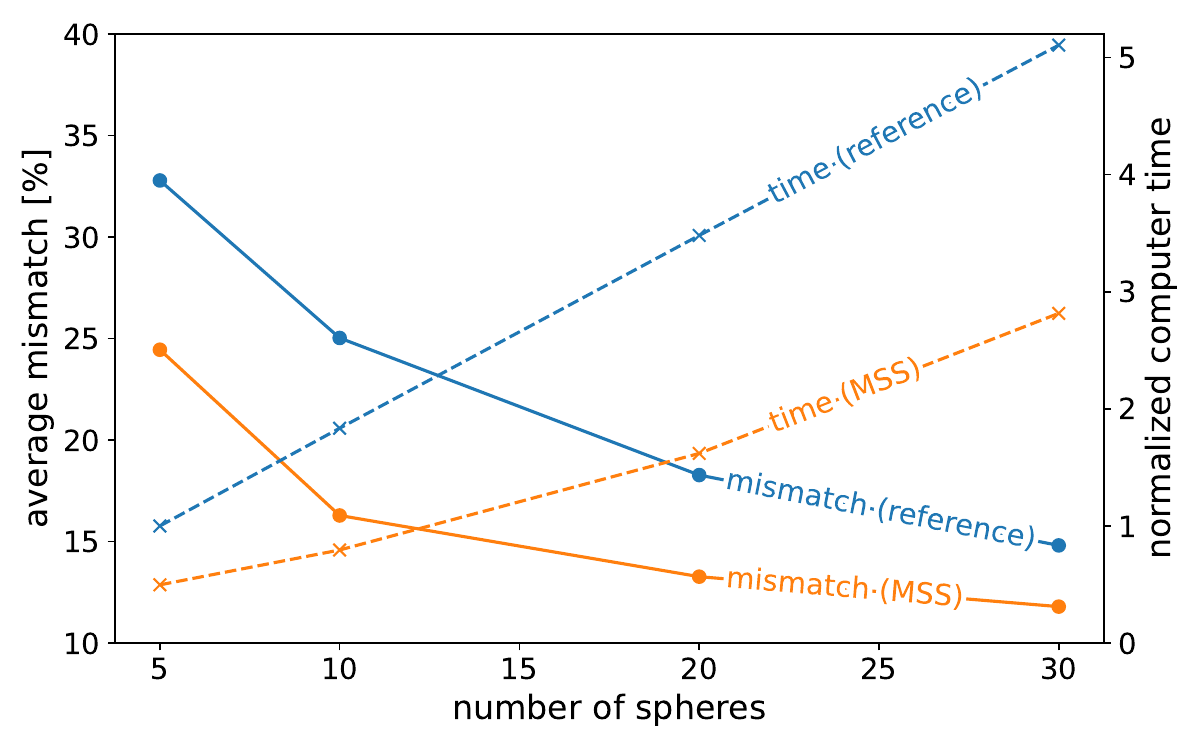}
    \caption{Average relative shape mismatch, $D$, for the multi-sphere representations of 402 particles obtained from the Sand Atlas \cite{Vego.2025}, together with the corresponding normalized computer time (scaled by the reference runtime at the lowest fidelity, $\approx$ 13 s). The results are shown as functions of the number of constituent spheres in $\tilde{\cal{S}}$.
    }
    \label{fig:sand-indicators}
\end{figure}

We applied MSS and the reference algorithm to 402 sand grains obtained from the public data base \textit{Sand Atlas} \cite{Vego.2025}. This database comprises a wide variety of representative particle shapes. Each target shape is voxelized on an axis-aligned bounding box using $205$ voxels along its smallest dimension; the voxel numbers along the remaining dimensions vary with the particle aspect ratio. \Cref{fig:sand-examples} 
shows examples of five of these target shapes and the multi-sphere representations generated by the reference algorithm and MSS. \Cref{fig:sand-indicators} 
shows the average relative shape mismatch, $D$, \Cref{eq:mismatch-discrete} between, the target shapes and the multi-sphere representations generated by MSS and the reference algorithm, as a function of the number of constituent spheres, $n=5,\dots,30$. In all cases, MSS achieves a substantial reduction of the mismatch at lower computational cost (also shown in \Cref{fig:sand-indicators}) compared to the reference algorithm.

\section{Summary}

We introduced the multi-sphere shape generator (MSS), a robust and efficient method for approximating arbitrary target geometries by means of overlapping spheres. The performance of MSS was systematically evaluated against \textsc{Clump} \cite{Angelidakis.2021} which is the best-performing algorithm available at the time.

Systematic benchmarks covering geometries of different complexity demonstrate that MSS consistently achieves smaller shape mismatches at substantially reduced computational cost, independent of the specific target geometry and the number of spheres employed. For comparison, we considered two benchmark shapes proposed in \cite{Angelidakis.2021}, using parameter sets optimized to ensure best possible performance of the reference algorithm. In contrast, MSS does not rely on user-defined parameters beyond the termination criterion. This represents a further significant advantage, as suitable parameter values for alternative approaches are generally not known \textit{\`a priori} and must be determined empirically for each class of particle shapes. The performance of MSS relative to the reference algorithm was further evaluated using a large variety of particle shapes from the Sand Atlas database \cite{Vego.2025}. These shapes are particularly relevant for DEM simulations of technical and geomechanical granular materials, in which individual grains are represented by multi-sphere particles.

Beyond computational performance, MSS exhibits important qualitative advantages. The algorithm preserves inherent symmetries of the target shape and avoids the generation of spurious spheres that would otherwise introduce artificial surface roughness. As a result, MSS provides not only more accurate but also physically more consistent multi-sphere representations.

MSS has been implemented in both \texttt{Python} and \texttt{C++} and is publicly available as open-source software \cite{Buchele.2026, Moradian.2026}. The algorithm can be integrated into DEM codes such as MercuryDPM, enabling its use within large-scale DEM simulations. The ability to generate new particle shapes at runtime—or to dynamically modify the shapes of existing ones—combined with the substantial gains in computational efficiency, makes MSS particularly well suited for advanced DEM applications involving plastic deformation and particle fragmentation \cite{Poeschel.2000}. In such scenarios, where particle geometries evolve during simulation, MSS enables accurate and computationally feasible multi-sphere representations without additional parameter tuning.

\section*{Acknowledgements}
Patric Müller acknowledges funding by Deutsche For\-schungs\-ge\-mein\-schaft (German Research Foundation) through research grant 398618334. 

Thorsten Pöschel acknowledges funding by Deutsche For\-schungs\-ge\-mein\-schaft (German Research Foundation) through research grants 446317499 and 377472739.

\appendix
\section{Optimal parameters used for the reference algorithm}
\label{app:parameters}
The reference algorithm \textsc{Clump} requires user-defined parameters, namely $div$, $rMin$, and $overlap$, which must be specified for each application. \Cref{tab:param}
\begin{table}[h!]
\centering
\begin{tabular}{lccc}
\toprule
shape & $div$ & $rMin$ & $overlap$ \\ \midrule
ellipsoid & $100$ & $0.25$ & $0.95$ \\
human femur bone & $150$ & $4$ & $0.95$\\
SAND atlas shapes & $200$& varying & $0.95$ \\
\bottomrule
\end{tabular}
\caption{Optimal parameters used for the reference algorithm  \cite{Angelidakis.2021}}
\label{tab:param} 
\end{table}
lists the optimal parameter sets that yielded the smallest mismatch, as defined by \Cref{eq:mismatch-dice}, for the benchmark cases described in \Cref{sec:performance}. These parameter sets were identified through systematic parameter sweeps and subsequently used for the comparison against MSS. For the physical interpretation of the user-defined parameters, see Ref. \cite{Angelidakis.2021}

\bibliographystyle{elsarticle-num}
\bibliography{shape-generator-references}

@article{Buchele.2026,
    author = {Buchele, Felix and M\"uller, Patric and Moradian, Arash and P\"oschel, Thorsten},
    title = {multisphere: a {P}ython implementation of the {MSS}-multi-shere shape generator for {DEM} simulations},
    journal = {preprint, available upon request},
    year = {2026}
}

@article{Moradian.2026,
    author = {Moradian, Arash and Buchele, Felix and M\"uller, Patric and P\"oschel, Thorsten},
    title = {The {MSS}-multi-sphere shape generator {C}++ library for integration in {DEM}},
    journal = {preprint, available upon request},
    year = {2026}
}

@article{Berry:2023,
title = {Contact models for the multi-sphere discrete element method},
journal = {Powder Technology},
volume = {416},
pages = {118209},
year = {2023},
doi = {j.powtec.2022.118209},
author = {Nathan Berry and Yonghao Zhang and Sina Haeri},
}

@article{Danielsson:1980,
  title={Euclidean distance mapping},
  author={Danielsson, Per-Erik},
  journal={Computer Graphics and Image Processing},
  volume={14},
  pages={227--248},
  year={1980},
  doi={10.1016/0146-664X(80)90054-4}
}

@article{Rosenfeld:1966,
  title={Sequential operations in digital picture processing},
  author={Rosenfeld, Azriel and Pfaltz, John L.},
  journal={Journal of the ACM (JACM)},
  volume={13},
  pages={471--494},
  year={1966},
  doi={10.1145/321356.321357}
}

@incollection{Meijster:2000,
author="Meijster, A. and Roerdink, J. B. T. M. and Hesselink, W. H.",
editor="Goutsias, John and Vincent, Luc and Bloomberg, Dan S.",
title="A General Algorithm for Computing Distance Transforms in Linear Time",
bookTitle="Mathematical Morphology and its Applications to Image and Signal Processing",
year="2000",
publisher="Springer US",
address="Boston, MA",
pages="331--340",
doi="10.1007/0-306-47025-X_36"
}

@article{Yuan:2019,
	author = {Yuan, Fei-Liang},
	journal = {Granular Matter},
	pages = {19},
	title = {Combined 3D thinning and greedy algorithm to approximate realistic particles with corrected mechanical properties},
	volume = {21},
	year = {2019},
	doi = {10.1007/s10035-019-0874-x},
}

@article{hertz:1882,
  author    = {Heinrich Hertz},
  title     = {Ueber die {B}er\"uhrung fester elastischer {K}\"orper},
  journal   = {Journal für die reine und angewandte Mathematik},
  volume    = {92},
  pages     = {156--171},
  year      = {1882},
doi = {10.1515/crll.1882.92.156}
}

@inproceedings{Amberger:2012,
author = {Amberger, Stefan and Friedl, Michael and Goniva, Christoph and Pirker, Stefan and Kloss, Christoph},
editor = {Eberhardsteiner, Josef and Böhm, Helmut J. and Rammerstorfer, Franz G.},
booktitle = {Proceedings of the 6th European Congress on Computational Methods in Applied Sciences and Engineering (ECCOMAS 2012)},
year = {2012},
pages = {},
title = {Approximation of Objects by Spheres for Multisphere Simulations in {DEM}},
Xjournal = {ECCOMAS 2012 - European Congress on Computational Methods in Applied Sciences and Engineering},
}

@article{Hubbard:1996, 
author = {Hubbard, Philip M.}, 
title = {Approximating polyhedra with spheres for time-critical collision detection}, 
year = {1996}, 
volume = {15}, 
doi = {10.1145/231731.231732}, 
journal = {ACM Transactions on Graphics}, 
pages = {179–210}, 
}

@article{Favier:1999,
author = {Favier, John and Abbaspour-Fard, Mohammad and Kremmer, M. and Raji, A. O.},
year = {1999},
pages = {467-480},
title = {Shape representation of axi-symmetrical, non-spherical particles in discrete element simulation using multi-element model particles},
volume = {16},
journal = {Engineering Computations},
doi = {10.1108/02644409910271894}
}

@article{BuchholtzPoeschel:1994,
author = {Buchholtz, Volkhard and Pöschel, Thorsten},
 doi = {10.1016/0378-4371(94)90467-7},
 journal = {Physica A-Statistical Mechanics and Its Applications},
 pages = {390-401},
 title = {Numerical investigations of the evolution of sandpiles},
 volume = {202},
 year = {1994},
}

@article{BuchholtzPoeschel:1996,
    author = {Buchholtz, V. and P\"oschel, T.},
    title = {Avalanche statistics of sand heaps},
    journal = {Journal of Statistical Physics},
volume = {84},
pages = {1373–1378},
    year = {1996},
doi = {10.1007/BF02174136},
}

@article{Cundall:1979,
    author = {Cundall, P. A. and Strack, O. D. L.},
    title = {A discrete numerical model for granular assemblies},
doi={10.1680/geot.1979.29.1.47},
    journal = {G\'eotechnique},
volume = {79},
pages = {47-65},
    year = {1979},
}

@book{Matuttis:2014,
    author = {Matuttis, Hans-Georg and Chen, Jian},
    title = {Understanding the Discrete Element Method: Simulation of Non-Spherical Particles for Granular and Multi-body Systems},
    publisher = {Wiley},
doi={10.1002/9781118567210},
    year = {2014},
}

@article{Silin.2006,
title = {Pore space morphology analysis using maximal inscribed spheres},
journal = {Physica A: Statistical Mechanics and its Applications},
volume = {371},
pages = {336-360},
year = {2006},
doi = {10.1016/j.physa.2006.04.048},
author = {Silin, Dmitriy and  Patzek, Tad}
}

@article{Gostick.2017,
title = {Versatile and efficient pore network extraction method using marker-based watershed segmentation},
author = {Gostick, Jeff T.},
journal = {Physical Review E},
volume = {96},
pages = {023307},
year = {2017},
doi = {10.1103/PhysRevE.96.023307}
}

@article{Khan.2019,
title = {Dual network extraction algorithm to investigate multiple transport processes in porous materials: {I}mage-based modeling of pore and grain scale processes},
journal = {Computers \& Chemical Engineering},
volume = {123},
pages = {64-77},
year = {2019},
doi = {10.1016/j.compchemeng.2018.12.025},
author = {Khan, Zohaib Atiq  and Tranter, Tom  and Agnaou, Mehrez  and Elkamel, Ali  and Gostick, Jeff }
}

@article{Angelidakis.2021,
title = {{CLUMP}: {A} Code Library to generate Universal Multi-sphere Particles},
journal = {SoftwareX},
volume = {15},
pages = {100735},
year = {2021},
doi = {10.1016/j.softx.2021.100735},
author = {Angelidakis, Vasileios and  Nadimi, Sadegh and Otsubo, Masahide and Utili, Stefano}
}

@article{Canbolat.2025,
title = {A {P}ython implementation of {CLUMP}, the Code Library to generate Universal Multi-sphere Particles},
journal = {SoftwareX},
volume = {29},
pages = {101957},
year = {2025},
doi = {10.1016/j.softx.2024.101957},
author = {Canbolat, Ahmet Utku and Nadimi, Sadegh  and Angelidakis, Vasileios}
}

@article{Sezgin.2004,
author = {Sezgin, Mehmet and Sankur, B{\"u}lent},
title = {{Survey over image thresholding techniques and quantitative performance evaluation}},
volume = {13},
journal = {Journal of Electronic Imaging},
pages = {146 -- 165},
year = {2004},
doi = {10.1117/1.1631315}
}

@ARTICLE{Otsu.1979,
author={Nobuyuki Otsu},
journal={IEEE Transactions on Systems, Man, and Cybernetics}, 
title={A Threshold Selection Method from Gray-Level Histograms}, 
year={1979},
volume={9},
pages={62-66},
doi={10.1109/TSMC.1979.4310076}}

@article{Zhao.2023,
title = {The role of particle shape in computational modelling of granular matter},
journal = {Nature Reviews Physics},
volume = {5},
year = {2023},
doi = {10.1038/s42254-023-00617-9},
author = {Jidong Zhao and Shiwei Zhao and Stefan Luding}
}

@article{Feng.2023,
  title = {Thirty years of developments in contact modelling of non-spherical particles in {DEM}: a selective review},
  author = {Feng, Y. T.},
  journal = {Acta Mechanica Sinica},
  volume = {9},
  year = {2024},
  doi = {10.1007/s10409-022-22343-x}
}

@article{Poeschel.1993,
  title = {Static friction phenomena in granular materials: {C}oulomb law versus particle geometry},
  author = {P\"oschel, Thorsten and Buchholtz, Volkhard},
  journal = {Physical Review Letters},
  volume = {71},
  pages = {3963--3966},
  year = {1993},
  doi = {10.1103/PhysRevLett.71.3963}
}

@article{Poeschel.2000,
  title = {Molecular dynamics of comminution in ball mills},
  author = {Buchholtz, V. and Freund, J. A. and P\"oschel, T.},
  journal = {European Physical Journal B},
  volume = {16},
  pages = {1434-6036},
  year = {2000},
  doi = {10.1007/PL00011052}
}

@book{Poeschel.2005,
  author    = {Thorsten Pöschel and Thomas Schwager},
  title     = {Computational Granular Dynamics: Models and Algorithms},
  publisher = {Springer},
  address   = {Berlin, Heidelberg},
  year      = {2005},
  doi       = {10.1007/3-540-27720-X}
}

@article{Bradshaw.2004,
  title={Adaptive medial-axis approximation for sphere-tree construction},
  author={Bradshaw, Gareth and O'Sullivan, Carol},
  journal={ACM Transactions on Graphics (TOG)},
  volume={23},
  pages={1--26},
  year={2004},
doi={10.1145/966131.966132}
}

@article{Price.2007,
  title={Sphere clump generation and trajectory comparison for real particles},
  author={Price, Mathew and Murariu, Vasile and Morrison, Garry},
  journal={Proceedings of discrete element modelling 2007},
  pages={105--113},
  year={2007},
xdoi={keine DOI}
}

@article{Wang.2007,
  title={Representation of real particles for {DEM} simulation using {X}-ray tomography},
  author={Wang, Linbing and Park, Jin-Young and Fu, Yanrong},
  journal={Construction and Building Materials},
  volume={21},
  pages={338--346},
  year={2007},
doi={10.1016/j.conbuildmat.2005.08.013}
}

@article{Garcia.2009,
  title={A clustered overlapping sphere algorithm to represent real particles in discrete element modelling},
  author={Garcia, X and Latham, J-P and Xiang, Jin-sheng and Harrison, JP},
  journal={G{\'e}otechnique},
  volume={59},
  pages={779--784},
  year={2009},
doi={10.1680/geot.8.T.037}
 }

@article{Ferellec.2010,
  title={A method to model realistic particle shape and inertia in {DEM}},
  author={Ferellec, Jean-Francois and McDowell, Glenn R},
  journal={Granular Matter},
  volume={12},
  pages={459--467},
  year={2010},
  doi={10.1007/s10035-010-0205-8}
}

@inproceedings{Taghavi.2011,
  title={Automatic clump generation based on mid-surface},
  author={Taghavi, R},
  booktitle={Proceedings, 2nd international FLAC/DEM symposium, Melbourne},
  pages={791--797},
  year={2011},
xdoi={kein DOI}
}

@article{Markauskas.2010,
  title={Investigation of adequacy of multi-sphere approximation of elliptical particles for {DEM} simulations},
  author={Markauskas, Darius and Ka{\v{c}}ianauskas, Rimantas and D{\v{z}}iugys, Algis and Navakas, Robertas},
  journal={Granular Matter},
  volume={12},
  pages={107--123},
  year={2010},
  doi={10.1007/s10035-009-0158-y}
}

@article{Tamadondar.2020,
  title={The effect of carrier surface roughness on wall collision-induced detachment of micronized pharmaceutical particles},
  author={Tamadondar, Mohammad R and Rasmuson, Anders},
  journal={AIChE Journal},
  volume={66},
  pages={e16771},
  year={2020},
doi={10.1002/aic.16771}
}

@article{Nan.2017,
  title={Analysis of powder rheometry of {FT4}: {E}ffect of particle shape},
  author={Nan, Wenguang and Ghadiri, Mojtaba and Wang, Yueshe},
  journal={Chemical Engineering Science},
  volume={173},
  pages={374--383},
  year={2017},
doi={10.1016/j.ces.2017.08.004}
}

@article{Pasha.2016,
  title={Effect of particle shape on flow in discrete element method simulation of a rotary batch seed coater},
  author={Pasha, Mehrdad and Hare, Colin and Ghadiri, Mojtaba and Gunadi, Alfeno and Piccione, Patrick M},
  journal={Powder Technology},
  volume={296},
  pages={29--36},
  year={2016},
  doi={10.1016/j.powtec.2015.10.055}
}

@article{Cho.2007,
  title={A clumped particle model for rock},
  author={Cho, N al and Martin, CD and Sego, DC},
  journal={International journal of rock mechanics and mining sciences},
  volume={44},
  pages={997--1010},
  year={2007},
doi={10.1016/j.ijrmms.2007.02.002}
}

@article{Kodicherla.2020,
  title={Investigations of the effects of particle morphology on granular material behaviors using a multi-sphere approach},
  author={Kodicherla, Shiva Prashanth Kumar and Gong, Guobin and Fan, Lei and Wilkinson, Stephen and Moy, Charles KS},
  journal={Journal of Rock Mechanics and Geotechnical Engineering},
  volume={12},
  pages={1301--1312},
  year={2020},
doi={10.1016/j.jrmge.2020.04.005}
}

@article{Gao.2012,
  title={A new method to simulate irregular particles by discrete element method},
  author={Gao, Rui and Du, Xin and Zeng, Yawu and Li, Yong and Yan, Jing},
  journal={Journal of Rock Mechanics and Geotechnical Engineering},
  volume={4},
  pages={276--281},
  year={2012},
doi={10.3724/SP.J.1235.2012.00276}
}

@article{Li.2015,
  title={Multi-sphere approximation of real particles for {DEM} simulation based on a modified greedy heuristic algorithm},
  author={Li, Cheng-Qing and Xu, Wen-Jie and Meng, Qing-Shan},
  journal={Powder Technology},
  volume={286},
  pages={478--487},
  year={2015},
  doi={10.1016/j.powtec.2015.08.026}
}

@article{Zheng.2016,
  title={Roundness and sphericity of soil particles in assemblies by computational geometry},
  author={Zheng, Junxing and Hryciw, Roman D},
  journal={Journal of Computing in Civil Engineering},
  volume={30},
  pages={04016021},
  year={2016},
doi={10.1061/(ASCE)CP.1943-5487.0000578}
}

@article{Haeri.2017,
  title={Optimisation of blade type spreaders for powder bed preparation in Additive Manufacturing using {DEM} simulations},
  author={Haeri, Sina},
  journal={Powder Technology},
  volume={321},
  pages={94--104},
  year={2017},
  doi={10.1016/j.powtec.2017.08.011}
}

@article{Katagiri.2019,
  title={A novel way to determine number of spheres in clump-type particle-shape approximation in discrete-element modelling},
  author={Katagiri, Jun},
  journal={G{\'e}otechnique},
  volume={69},
  pages={620--626},
  year={2019},
doi={10.1680/jgeot.18.P.021}
}

@article{Favier.2001,
author = {J. F. Favier  and M. H. Abbaspour-Fard  and M. Kremmer },
title = {Modeling Nonspherical Particles Using Multisphere Discrete Elements},
journal = {Journal of Engineering Mechanics},
volume = {127},
pages = {971-977},
year = {2001},
doi = {10.1061/(ASCE)0733-9399(2001)127:10(971)}
}

@ARTICLE{Maurer.2003, 
author={Maurer, C. R. and Rensheng Qi and Raghavan, V.}, 
journal={IEEE Transactions on Pattern Analysis and Machine Intelligence}, 
title={A linear time algorithm for computing exact Euclidean distance transforms of binary images in arbitrary dimensions}, 
year={2003}, 
volume={25}, 
pages={265-270},
doi={10.1109/TPAMI.2003.1177156}
}

@ARTICLE{Vego.2025, 
author={Vego, I. and Marks, B.}, 
journal={Geoscience Data Journal}, 
title={The Sand Atlas}, 
year={2025}, 
volume={12}, 
doi={10.1002/gdj3.70008}
}

@ARTICLE{Luijmes.2024,
author={Luijmes, Joost
and van Leeuwen, Tristan
and Renema, Willem},
title={ForametCeTera, a novel CT scan dataset to expedite classification research of (non-)foraminifera},
journal={Scientific Data},
year={2024},
volume={11},
pages={642},
doi={10.1038/s41597-024-03476-w}
}

@phdthesis{Vego.2023,
    author = {Vego, Ilija},
    title = {Multi-modal investigation of hygroscopic granular media at high relative humidity},
    school = {Université Grenoble Alpes},
    year = {2023}
}

@INPROCEEDINGS{Saadatfar.2012,
  title      = "{3D} mapping of deformation in an unconsolidated sand: {A} micro mechanical study",
  booktitle  = "{SEG} Technical Program Expanded Abstracts 2012",
  author     = "Saadatfar, Mohammad and Francois, Nicolas and Arad, Alon and Madadi, Mahyar and Cruikshank, Ron and Alizadeh, Mehdi and Sheppard, Adrian and Kingston, Andrew and Limay, Ajay and Senden, Tim and Knackstedt, Mark",
  doi={10.1190/segam2012-1263.1},
  publisher  = "Society of Exploration Geophysicists",
  month      =  sep,
  year       =  2012,
pages={1-6},
\doi={10.1190/segam2012-1263.1},
  conference = "SEG Technical Program Expanded Abstracts 2012"
}

@ARTICLE{Wiebicke.2017,
  title     = "On the metrology of interparticle contacts in sand from x-ray tomography images",
  author    = "Wiebicke, Max and And{\`o}, Edward and Herle, Ivo and Viggiani, Gioacchino",
  journal   = "Meas. Sci. Technol.",
  publisher = "IOP Publishing",
  volume    =  28,
  number    =  12,
  pages     = "124007",
  month     =  dec,
  year      =  2017,
doi={10.1088/1361-6501/aa8dbf}
}

@Article{Milatz2021,
author={Milatz, Marius
and H{\"u}sener, Nicole
and And{\`o}, Edward
and Viggiani, Gioacchino
and Grabe, J{\"u}rgen},
title={Quantitative {3D} imaging of partially saturated granular materials under uniaxial compression},
journal={Acta Geotechnica},
year={2021},
volume={16},
pages={3573-3600},
doi={10.1007/s11440-021-01315-5}
}

@article{Dice.1945,
author = {Dice, Lee R.},
title = {Measures of the Amount of Ecologic Association Between Species},
journal = {Ecology},
volume = {26},
number = {3},
pages = {297-302},
doi = {https://doi.org/10.2307/1932409},
year = {1945}
}

@Article{Levy.2025,
author={Levy, Avivit
and Shalom, B. Riva
and Chalamish, Michal},
title={A guide to similarity measures and their data science applications},
journal={Journal of Big Data},
year={2025},
month={Jul},
day={26},
volume={12},
number={1},
pages={188},
issn={2196-1115},
doi={10.1186/s40537-025-01227-1}
}
\end{document}